\documentclass[lettersize,journal]{IEEEtran}
\usepackage{amsmath,amsfonts}
\usepackage{algorithmic}
\usepackage{algorithm}
\usepackage{array}
\usepackage[caption=false,font=normalsize,labelfont=sf,textfont=sf]{subfig}
\usepackage{textcomp}
\usepackage{stfloats}
\usepackage{url}
\usepackage{verbatim}
\usepackage{graphicx}
\usepackage{cite}
\usepackage{amsmath}
\usepackage{optidef}

\usepackage{tikz}
\usetikzlibrary{shapes, arrows, positioning, matrix, calc, fit}

\allowdisplaybreaks

\begin{document}

\title{A Model Predictive Control Approach to Dual-Axis Agrivoltaic Panel Tracking
}

\author{Anna Stuhlmacher \IEEEmembership{Member, IEEE},  Panupong Srisuthankul, Johanna L. Mathieu \IEEEmembership{Senior Member, IEEE}, and Peter Seiler  \IEEEmembership{Fellow, IEEE}\thanks{A. Stuhlmacher is with Michigan Technological University, Houghton, MI (email: annastu@mtu.edu). P. Srisuthankul, J.L. Mathieu, and P. Seiler are with the University of Michigan, Ann Arbor, MI (email: \{panusri, jlmath, pseiler\}@umich.edu).}}

\maketitle

\begin{abstract}
Agrivoltaic systems--photovoltaic (PV) panels installed above agricultural land--have emerged as a promising dual-use solution to address competing land demands for food and energy production. In this paper, we propose a model predictive control (MPC) approach to dual-axis agrivoltaic panel tracking control that dynamically adjusts panel positions in real time to maximize power production and crop yield given solar irradiance and ambient temperature measurements. We apply convex relaxations and shading factor approximations to reformulate the MPC optimization problem as a convex second-order cone program that determines the PV panel position adjustments away from the sun-tracking trajectory. Through case studies, we demonstrate our approach, exploring the Pareto front between i) an approach that maximizes power production without considering crop needs and ii) crop yield with no agrivoltaics. We also conduct a case study exploring the impact of forecast error on MPC performance. We find that dynamically adjusting agrivoltaic panel position helps us actively manage the trade-offs between power production and crop yield, and that active panel control enables the agrivoltaic system to achieve land equivalent ratio values of up to 1.897. 
\end{abstract}


\section{Introduction}\label{section: Intro}
\IEEEPARstart{I}{n} 2025, solar photovoltaic (PV) systems were the largest source of generation capacity added to the U.S. electric power grid for the fifth consecutive year~\cite{SEIA_2026}. A challenge associated with large-scale PV deployment is that it often competes for land that is also well suited for agriculture~\cite{turnley2024viability}. Agrivoltaics has emerged as a promising approach to address these competing land use demands of energy and food production~\cite{dinesh_2016, marrou_2019}. In agrivoltaics, elevated PV panels are installed over cropland, grazing areas, or pollinator habitats~\cite{macknick_2022}. Early experimental and modeling-based studies suggest that agrivoltaic systems can achieve higher overall land use efficiency than either agriculture or traditional ground-mounted PV installations alone~\cite{macknick_2022}. Furthermore, co-locating crops and PV panels may provide additional synergistic benefits, including improved PV efficiency from evaporative cooling~\cite{barron-gafford_2019, hassanpour2018remarkable}, reduced water irrigation needs due to crop shading~\cite{barron-gafford_2019,elamri_2018, hassanpour2018remarkable}, creation of favorable micro-climates supporting crop health~\cite{marrou_2013, hassanpour2018remarkable}, and, possibly, pest mitigation~\cite{thakur2025exploring}. 

The interdisciplinary field of agrivoltaics research is growing~\cite{mamun_review_2022, gomez2023knowns,widmer2024agrivoltaics,zheng_photovoltaics_2024} with research spanning engineering design~\cite{miljkovic2024advancing, campana_optimisation_2021}, crop/soil science~\cite{elamri_2018}, social science~\cite{pascaris_first_2020, pascaris_integrating_2021,moore_can_2022}, and policy~\cite{pascaris2021examining}. There are also a large number of new experimental and outreach-focused agrivoltaic installations~\cite{trommsdorff_2021, purdue2023,uiuc2023,jacks2023}. As interest in agrivoltaic systems grows, it is increasingly important to understand how agrivoltaic systems can impact the power grid. These impacts are highly dependent on agricultural choices, PV system design and operation choices, and climate. Engineering research on agrivoltaics has primarily focused on the impact of PV system design, not the impact of PV operation choices, despite the fact that panel tracking algorithms can significantly impact power production and crop health/productivity~\cite{hussain2024evaluating,alam2024does}. Agrivoltaic panel tracking schemes generally aim to (approximately) maximize power capture and typically employ standard PV tracking approaches using predetermined schedules based on location, time of year, etc. Feedback control, leveraging real-time meteorological and crop system measurements, has significant potential to improve the performance of agrivoltaic systems. In this paper, we propose a model predictive control (MPC) approach to optimize the real-time operational control of dual-axis PV panels within an agrivoltaic system. The objective is to dynamically adjust PV panel position to maximize the combined value of PV energy production and crop yield. 

A growing body of research has examined the performance, cost/benefits, and broader implications of agrivoltaic systems, usually on a case-by-case basis. Several studies use modeling and simulation to evaluate the effects of different panel configurations and management strategies~\cite{amaducci2018,dupraz_combining_2011, WuNAPS2025}. An experimental testbed is developed in \cite{barron-gafford_2019} that measures an agrivoltaic system's crop yield, environmental conditions, and PV power output across multiple crops. Refs.~\cite{imran2020optimization,gupta2024optimizing, GRUBBS2024114018, abidin,mazzeo2025optimizing} develop modeling approaches to capture crop-PV interactions and run simulations to find ``optimal" designs/tracking
approaches, but do not formulate explicit optimization problems. These studies highlight the potential benefits and trade-offs in agrivoltaic systems; however,  formal optimization and control approaches are important tools to systematically identify the best design and operation choices. 

Very few studies develop agrivoltaic system design or operation optimization formulations. We briefly review three such papers. First, \cite{campana_optimisation_2021} applies a genetic algorithm to solve for PV design parameters (azimuth and row distance) of static bifacial vertically mounted agrivoltaic panels. Second, \cite{miljkovic2024advancing} applies a genetic algorithm to solve for PV design parameters (row spacing, pole height, and crop buffer zone) of east-west single-axis tracking bifacial agrivoltaic panels, with the tracking algorithm designed to maximize power capture. Third, \cite{fu2022collaborative} applies particle swarm optimization to optimize the geometric arrangement of PV panels on greenhouse rooftops (and associated greenhouse load controls). 

To the best of our knowledge, no agrivoltaic research, other than our own preliminary
work~\cite{StuhlmacherHICSS2024}, recent work by Fraunhofer~\cite{bruno2025enhancing}, and recent work by Mignoni et al.~\cite{Mignoni2025, Mignoni2025b} has developed approaches to actively and dynamically control, in real time, single- or dual-axis agrivoltaic panels considering both power capture and crop needs. Ref.~\cite{bruno2025enhancing} dynamically optimizes single-axis tracking to meet light requirements for crops, though it uses a heuristic optimization approach (a simulation and Fibonacci Search Algorithm) rather than a formal optimal control approach. Refs.~\cite{Mignoni2025, Mignoni2025b} formulate a non-convex
mixed-integer optimization problem and approximate it with a convex mixed-integer formulation for a single-axis tracking problem that balances power production and shading, assuming an ideal shading trajectory; however, it is not clear how to determine that trajectory since the approach does not include a crop model. In contrast, we formulate a dual-axis agrivoltaic panel tracking control problem as an MPC problem subject to the PV system and crop constraints. 

In our preliminary work~\cite{StuhlmacherHICSS2024}, we optimally solved for the daily operation of a dual-axis tracking agrivoltaic system, where we ensured that the field below the PV panels receives a prescribed minimum amount of daily intercepted photosynthetically active radiation (PAR), which is the range of solar radiation wavelengths useful for plant photosynthesis. We assumed that we knew the target daily PAR threshold in advance (i.e., based on the plant's light saturation point) and we did not explicitly model crop outcomes. In this paper, we significantly extend that work in three ways. First, we consider the operation over the entire growing season to explicitly evaluate the trade-offs between crops and PV panels. Second, we incorporate a crop growth model to capture crop yield subject to climate, intercepted PAR, and crop physiological traits. And third, we adopt an MPC framework to account for irradiance and temperature forecast uncertainty, enabling dynamic control decisions based on updated forecasts.

The contributions of this paper are 1) development of an MPC approach to agrivoltaic panel tracking control to optimize PV power production and crop yield, including formulation of the MPC optimization problem with models/constraints capturing how panel position impacts crop shading and how crop shading and temperature impacts crop yield; 2) reformulation of the optimization problem into a convex second order cone program (SOCP) using trigonometric properties and shading factor approximations; and 3) case studies demonstrating the performance of the approach. Specifically, we identify the Pareto front between i) an approach that maximizes power production without considering crop needs and ii) crop yield without agrivoltaics. We also explore the impact of forecast error on MPC performance. 

The rest of the paper is organized as follows. Section~\ref{section: overview} provides an overview of the MPC framework. Section~\ref{section: Model} presents the agrivoltaic constraints, while Section~\ref{section: Optimization} formulates the optimization problem and reformulates it as an SOCP. We demonstrate our approach through case studies in Section~\ref{section: Case Study}, and a discussion and concluding remarks are provided in Sections~\ref{section: Discussion} and \ref{section: Conclusion}.

\section{Problem overview} \label{section: overview}
We consider an agrivoltaic system consisting of crops grown beneath dual-axis tracking PV panels. Compared to fixed-tilt PV panels, agrivoltaic systems with dual- or single-axis tracking PV panels are able to adjust the PV panels dynamically throughout the day to allow the system to significantly increase power production and/or allow more sunlight to reach the crops below. In this paper, we focus on dual-axis agrivoltaic panels; however, the proposed formulation can also be simplified to single-axis configurations.  

We formulate the MPC optimization problem as a multi-period problem over a growing season, where the time periods are linked by crop dynamics, specifically biomass accumulation leading to end-of-season crop yield. Our control decisions are dynamic agrivoltaic panel adjustments that balance power production and crop growth. In each time period (e.g., every hour), the problem is solved with updated forecasts for solar irradiance and ambient temperature, generating control decisions for each time period, but only the first decision is implemented, following a standard MPC approach. 

One could use a hierarchical approach to decouple the growing season problem from the daily operational problem to reduce computational burden. For example, in~\cite{StuhlmacherHICSS2024}, we defined the growing season problem as the problem of choosing daily thresholds for intercepted PAR. We defined the daily operational problem as the problem of controlling the system to surpass the threshold PAR for that day. However, this approach is suboptimal as it does not explicitly capture crop outcomes and we have found that, through convex relaxations and approximations, our optimization formulation is sufficiently tractable to enable consideration of the full (or rest of the) growing season in the optimization problem solved over all time steps, and so we do not use this hierarchical approach here. 
 
\begin{figure}[t]
  \centering
  \resizebox{\columnwidth}{!}{%
    \begin{tikzpicture}[
      font=\small,
      >=stealth,
      block/.style = {rectangle, draw,  minimum height=1cm, align=center},
      smallblock/.style = {rectangle, draw,  minimum height=0.9cm, align=center},
      dottedbox/.style = {rectangle, draw, dashed, inner sep=6pt},
    ]
    \node[block, minimum width=2.5cm] (mpc) {MPC};
    
    \node[smallblock, right=2.2cm of mpc, minimum width = 1.3cm, yshift=0.2cm] (pv) {PV\\ Array};
    \node[smallblock, below=0.3cm of pv, minimum width = 1.3cm] (crops) {Crops};
    
    \node[dottedbox, fit={(pv)(crops)}] (ag) {};
    \node[below=0.05cm of ag, align=center] {Agrivoltaic \\System};
    
    \node[block, right=2.3cm of crops] (bio) {Biomass\\Calc};
    \node[block, right=0.5cm of bio] (yield) {Yield\\Calc};
    
    \node[above=0.7cm of mpc, xshift=-1cm, align=center] (solar)
    {\textit{Sun}\\\textit{Position}\\$\phi_\text{s}^{t},\,\beta_\text{s}^{t}$};
    
    \node[above=0.7cm of mpc, xshift=2cm, align=center] (irr)
    {\textit{Real-Time Measurements }\\\textit{and Forecasts: Temperature $T^{t}$}\\\textit{and Irradiance}
    $\text{DHI}^{t},\,\text{DNI}^{t}$};
    
    \coordinate (mpcSolarTop) at ([xshift=-1cm]mpc.north);
    \coordinate (mpcIrrTop)   at ([xshift= 2cm]mpc.north);
    
    \draw[->] ([xshift=0.2cm]solar.south) -- ([xshift=0.2cm]mpcSolarTop);
    \draw[->] ([xshift=-1cm]irr.south)   -- ([xshift=-1cm]mpcIrrTop);
    
    \draw[->] ([yshift=0.2cm]mpc.east) -- node[below, pos=0.4, align=center]
    {\textit{PV Position}\\$\Sigma^{t}_\text{pv},\,\phi^{t}_\text{pv}$} (pv.west);
    
    \draw[->] ([xshift=1.8cm,yshift=0.2cm]mpc.east) |- (crops.west);
    
    \node[right=5cm of pv, align=left] (power)
    {\textit{Power}\\ \textit{Produced} $P^{t}$};
    
    \draw[->] (pv.east) -- (power.west);
    
    \draw[->] (crops.east) -- node[below, align=center]
    {\textit{PAR}\\\textit{Intercepted}\\$\text{PAR}^{d}_\text{crop}$ } (bio.west);
    
    \draw[->] (bio.east) -- (yield.west);
    \node[below=0.5cm of bio, xshift=2.5cm, align=center ] {\textit{Biomass $B^{d}$ and}\\ \textit{Leaf Area Index} $\text{LAI}^{d}$};
    
    \node[right=0.3cm of yield, align=left] (yout)
    {\textit{Crop Yield}\\$Y$};
    
    \draw[->] (yield.east) -- (yout.west);
    
    \draw[->] ([xshift=0.2cm]bio.east) |- ([yshift=-3cm, xshift=-0.2cm]mpc.west) |- (mpc.west);
    
    \end{tikzpicture}
  }
  \caption{Block diagram of MPC approach to agrivoltaic panel tracking.}
  \label{fig: block diagram}
\end{figure}
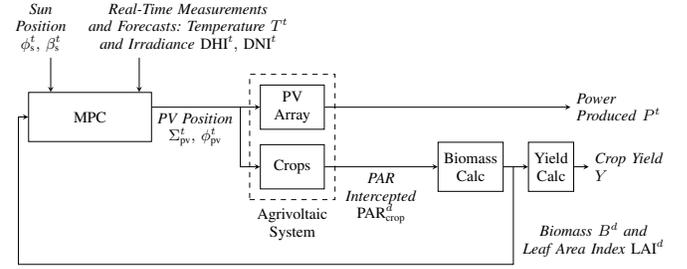

Fig.~\ref{fig: block diagram} shows a block diagram of our proposed controller. At each time step $t$, the MPC algorithm takes in the known sun position (solar azimuth $\phi_\mathrm{s}^t$ and altitude $ \beta_\mathrm{s}^t$), real-time measurements of solar irradiance (Diffuse Horizontal Irradiance $\mathrm{DHI}^t$, and Direct Normal Irradiance 
$\mathrm{DNI}^t$) and ambient temperature ($T^t$) along with forecasts of these values for the remainder of the control horizon, and estimates of the crop biomass ($B^d$) and leaf area index (LAI, denoted $\mathrm{LAI}^d$) from the end of the previous day $d$. It uses these values to compute the optimal PV panel position (PV tilt $\Sigma_\mathrm{pv}^t$ and azimuth $\phi_\mathrm{pv}^t$), specifically, deviations away from panel positions computed by a sun-tracking algorithm, i.e., panel positions that maximize power output from direct beam irradiance. These panel adjustments are applied to the agrivoltaic system, resulting in power production ($P^t$) and daily intercepted PAR ($\mathrm{PAR}_\mathrm{crop}^d$). Since crop growth is relatively slow, crop biomass and leaf area index are only estimated once per day. Specifically, at the end of each day, the daily intercepted PAR is used (together with the known solar irradiance and temperature measurements) to calculate the current crop biomass, temperature stress, and leaf area index (which is fed back to the controller). At the end of the growing season, crop biomass is used to estimate the end-of-season yield ($Y$), following the model in~\cite{singh}. 

We quantify the agrivoltaic operational value by assessing the land equivalent ratio (LER). LER (also referred to as land use efficiency) is the normalized value of crop yield and energy revenue relative to their respective single-use cases. This metric is commonly used in agrivoltaics to capture the combined productivity of dual-use land, where using the land for both can produce better efficiencies than using the land for one or the other (i.e., a cumulative LER of crops and PV is greater than one). We denote the total LER of the agrivoltaic system as $\text{LER}_\text{total}$
\begin{align}
    \text{LER}_\text{total} &= \text{LER}_\text{crop} + \text{LER}_\text{pv}, \label{eqn: LER total}
\end{align}
where $\text{LER}_\text{crop}$ and $\text{LER}_\text{pv}$ are the crop yield and PV revenue LERs. Here, $\text{LER}_\text{crop}$ is the ratio of the optimal agrivoltaic crop yield relative to the crop-only case without PV panels. Similarly, $\text{LER}_\text{pv}$ is the ratio of the optimal agrivoltaic energy revenue relative to the PV power-maximizing case. All modeling details are provided in the next section.

\section{Agrivoltaic model}\label{section: Model}
In this section, we present the agrivoltaic operational formulation. We solve for the PV panel adjustments away from the sun-tracking position to maximize the crop yield and PV energy revenue. We solve the problem for the entire growing season with $T$ time steps of duration $\Delta T$ (h), i.e., $t=1,...,T$  over all days within the growing season, i.e., $d=1,...,D$, where $D$ is the final day in the growing season. In the following subsections, the model inputs, variables, and constraints are defined. 

\subsection{Model inputs}
\begin{figure}
    \centering
	\includegraphics[width=\linewidth]{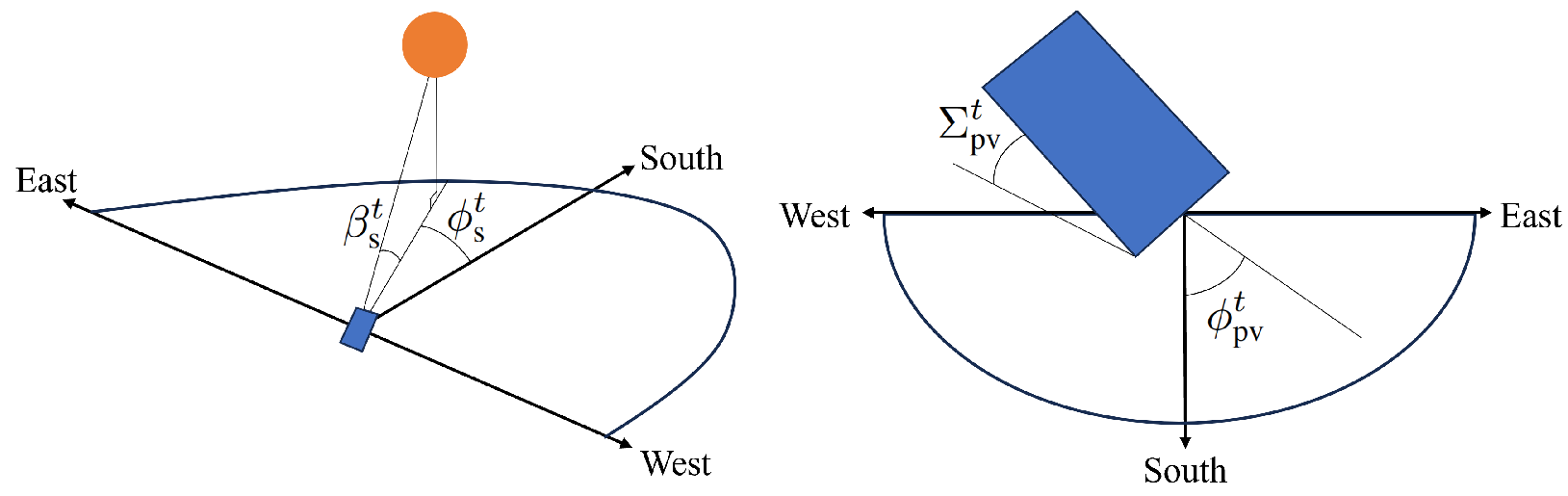}
	\caption{(Left) The sun's position can be characterized by the solar azimuth $\phi_\text{s}^t$ and altitude $\beta_\text{s}^t$. (Right) PV panel position can be characterized by the panel azimuth $\phi_\text{pv}^t$ and tilt $\Sigma_\text{pv}^t$.}
	\label{fig: diagram}       
\end{figure}

The operational problem depends on sun position, temperature,  and irradiance. Sun position is determined by geographic location and time, and is used for both sun tracking and crop shading calculations. At time $t$, the sun's position can be characterized by the solar azimuth angle $\phi_\text{s}^t$ ($^\circ$), defined as the angle between the sun and true south, and the solar altitude angle $\beta_\text{s}^t$ ($^\circ$), defined as the angle between the sun and the local horizon. These sun position angles are illustrated in the left diagram of Fig.~\ref{fig: diagram}. 

Both power production and crop growth depend on solar irradiance. Solar irradiance can be decomposed into direct beam, diffuse, and reflected components, where direct beam is usually the largest component. The diffuse component is particularly important in agrivoltaic systems since crops are partially shaded by the PV panels~\cite{MaLu_PAR}. Therefore, we consider the direct beam and diffuse irradiance components, denoted $I_\text{db}^t$ and $I_\text{diff}^t$ ($\text{W}/\text{m}^2$), respectively. Reflected irradiance is not considered in this formulation, as its effect on monofacial panels is usually modest~\cite{pulido2022spatial}. In our case studies, we use measurements of the Direct Normal Irradiance (DNI) and Diffuse Horizontal Irradiance (DHI) to determine the irradiance incident on the PV panels as well as the PAR received by the crops. The total available direct beam and diffuse PAR--denoted $\text{PAR}_\text{total,db}^t$ and $\text{PAR}_\text{total,diff}^t$ ($\text{W}/\text{m}^2$)--are estimated from DNI and DHI at time~$t$
\begin{align}
    \text{PAR}_\text{total,db}^t = \alpha \text{DNI}^t,\qquad  \text{PAR}_\text{total,diff}^t = \alpha \text{DHI}^t,\label{eqn: PAR GHI ratio}
\end{align}
where $\alpha$ is the PAR-irradiance ratio used in the EPIC crop model formulation~\cite{singh}. It should be noted that more detailed PAR decomposition methods (e.g.,~\cite{MaLu_PAR}) could also be employed since PAR is an input to our MPC formulation and does not affect the computational complexity of the optimization problem. 

We calculate the power produced by the PV panels given the irradiance and panel position in Section~\ref{subsection: PV power} and the PAR that reaches the field in Section~\ref{subsection: shading analysis}, taking field shading into account within our crop model in Section~\ref{subsection: crop model}.

\subsection{PV power output}\label{subsection: PV power}
The position of a PV panel at time $t$ can be characterized by the PV panel's azimuth $\phi_{\text{pv}}^t$ ($^\circ$), defined as the angle between the normal vector of the PV panel surface and true south (where east of south is positive), and tilt $\Sigma_\text{pv}^t$ ($^\circ$), defined as the angle between the PV collector surface and the horizontal plane. These PV panel position angles are illustrated in the right diagram of Fig.~\ref{fig: diagram}. We assume that all panels within an agrivoltaic system follow the same position controls since it is common for PV panels within a single installation to be operated the same way. However, the notation can be easily modified to allow independent orientation decisions for each panel, if desired. The feasible range of PV panel positions is bounded by the physical design limits of the PV system,
\begin{align}
     \underline{\phi}_\text{pv} &\leq  \phi_\text{pv}^t  \leq \overline{\phi}_\text{pv},\label{eqn: PV azimuith limit}\\
    \underline{\Sigma}_\text{pv} &\leq \Sigma_\text{pv}^t  \leq \overline{\Sigma}_\text{pv},\label{eqn: PV tilt limit}
\end{align}
where $\smash{\underline{\phi}_\text{pv}}$ and $\smash{\overline{\phi}_\text{pv}}$ are the minimum and maximum azimuth, and $\underline{\Sigma}_\text{pv}$ and $\overline{\Sigma}_\text{pv}$ are the minimum and maximum tilt.

To evaluate the PV power output, we calculate the angle of incidence $\theta_\text{s-pv}^t$ ($^\circ$), which is the angle between the sun's rays and the normal vector of the PV panel surface. This angle is determined by the relative positions of the sun and PV panel at time $t$~\cite{masters2023renewable},
\begin{align}
    \!\cos{\theta_\text{s-pv}^t} = \cos{\beta_\text{s}^t} &\cos{(\phi_\text{s}^t-\phi_\text{pv}^t)}\sin{\Sigma_\text{pv}^t}+ \sin{\beta_\text{s}^t}\cos{\Sigma_\text{pv}^t}. \label{eqn: incidence angle}
\end{align}
Using measurements of DNI and DHI, the direct beam irradiance $I_\text{db}^t$ and diffuse irradiance $I_\text{diff}^t$ incident on the PV collector at time $t$ is~\cite{masters2023renewable}
\begin{align}
    I_\text{db}^t & = \text{DNI}^t\cdot\cos{\theta_\text{s-pv}^t} ,\label{eqn: db irradiance}\\
    I_\text{diff}^t & = \text{DHI}^t\cdot\left(\frac{1+\cos{\Sigma_\text{pv}^t}}{2}\right).\label{eqn: diff irradiance}
\end{align}
The power output of the PV system at time $t$, denoted $P^t$ (W), is
\begin{align}
    P^t &= A_\text{array} \cdot \eta_\text{array} \cdot \left(I_\text{db}^t + I_\text{diff}^t\right), \label{eqn: PV power}
\end{align}
where $A_\text{array}$ is the total surface area of the PV array ($\text{m}^2$) and $\eta_\text{array}$ is the array efficiency. In this formulation, we do not consider shading between panels, since agrivoltaic systems typically have wider row spacing than conventional PV installations.  

Given the power produced by the PV panels, we can calculate the monetary value of the energy produced,
\begin{align}
    \text{revenue}(\boldsymbol{P}) := \sum_{t=1}^T \pi^t \cdot P^t  \cdot \Delta T, \label{eqn: PV cost}
\end{align}
where $\boldsymbol{P}=[P^t \,\forall\, t=1...T]$ and function $\text{revenue}(\cdot)$ is the amount of money earned from the PV installation given the input of power injected into the grid and $\pi^t$ is the price of electricity produced ($\$/\text{Wh}$).

We assume that the dual-axis PV system follows a baseline sun-tracking trajectory, where adjustments are made from this trajectory to satisfy crop PAR needs. A simple sun-tracking  strategy maximizes the direct beam irradiance (i.e., $\theta_\text{pv-s}^t=0$),
\begin{align}
    \Sigma_\text{pv,st}^t &= 90^\circ - \beta_\text{s}^t,\label{eqn: ST tilt}\\
    \phi_\text{pv,st}^t & = \phi_\text{s}^t , \label{eqn: ST azimuth}
\end{align}
where $\Sigma_\text{pv,st}^t$ and $\phi_\text{pv,st}^t$ are the PV tilt and azimuth position at time $t$ when following the sun-tracking trajectory. 
To make adjustments from the sun-tracking trajectory, we define the PV panel position as
\begin{align}
    \Sigma_\text{pv}^t &= \delta\Sigma_\text{pv}^t + \Sigma_\text{pv,st}^t,\label{eqn: tilt adjust}\\
    \phi_\text{pv}^t &= \delta \phi_\text{pv}^t + \phi_\text{pv,st}^t,\label{eqn: azimuth adjust}
\end{align}
where $\delta \Sigma_\text{pv}^t$ and $\delta\phi_\text{pv}^t$ represent adjustments to the PV tilt and azimuth angles relative to the sun-tracking position. Using the irradiance and power definitions \eqref{eqn: db irradiance}-\eqref{eqn: PV power} together with the PV panel position definitions \eqref{eqn: tilt adjust}-\eqref{eqn: azimuth adjust}, we can express the PV irradiance deviations $\delta I_\text{db}^t$ and $\delta I_\text{diff}^t$ as well as the resulting power output deviation $\delta P^t$ relative to the sun-tracking trajectory \eqref{eqn: ST tilt}-\eqref{eqn: ST azimuth} at time $t$ as
\begin{align}  
    \delta I_\text{db}^t & =\text{DNI}^t\cdot \left ( \cos{(\theta_\text{s-pv}^t)} -1\right), \label{eqn: db irradiance adjust}\\
    \delta I_\text{diff}^t &= \text{DHI}^t\cdot\left(\frac{\cos{(\Sigma_\text{pv}^t)}-\cos{(\Sigma_\text{pv,st}^t)}}{2}\right), \label{eqn: diff irradiance adjust}\\
    \delta P^t &= A_\text{array} \cdot \eta_\text{array} \cdot( \delta I_\text{db}^t + \delta I_\text{diff}^t).\label{eqn: power adjust}
\end{align}
The PV LER is then 
\begin{align}
        \text{LER}_\text{pv} &=  \frac{ \text{revenue}(\boldsymbol{P}_\text{st} + \boldsymbol{\delta P})}{\text{revenue}(\boldsymbol{P}_\text{st})}, \label{eqn: LER pv}
\end{align}
where $\boldsymbol{P}_\text{st}$ is the vector of power output under sun tracking, and $\boldsymbol{\delta P}$ is the vector of the power deviations given adjustments away from sun tracking.

\subsection{Shading analysis}\label{subsection: shading analysis}
To determine the PAR that reaches the field at time $t$, we consider the direct beam and diffuse PAR components, where the direct beam PAR reaching the field depends on the portion of the field that is shaded by PV panels, 
\begin{align}
    \!\!\!\text{PAR}_\text{field}^t\! = \!\left(1-S_F(\delta\phi_\text{pv}^t, \delta\Sigma_\text{pv}^t)\right)\! \cdot \!\text{PAR}_\text{total,db}^t \!+\! \text{PAR}_\text{total,diff}^t, \label{eqn: PAR field SF}
\end{align}
where shading factor function $S_F(\cdot)$ depends on the positions of the PV panels and the sun. The shading factor represents the fraction of the field shaded by the PV panels at a given time and is calculated using the geometric shading calculation procedure described in~\cite{StuhlmacherHICSS2024}. Unlike~\cite{StuhlmacherHICSS2024}, shading only impacts the direct component of PAR, which is an important distinction for agrivoltaic systems~\cite{MaLu_PAR}. This approach can also be extended to evaluate shading at smaller spatial resolutions, e.g., if we want to capture the impact of edge and interrow shading effects. 

\subsection{Crop model}\label{subsection: crop model}
We use the Environmental Policy Integrated Climate (EPIC) crop model~\cite{williams1989epic, singh}, which estimates crop yield and biomass based on climatic data, crop information, and intercepted field PAR. EPIC uses a single modeling framework with crop-specific parameters for different crops. There is extensive literature on improving the crop parameter accuracy, extending to additional crops, and adding additional components, with spin-off models and software packages, such as APEX and WinEPIC~\cite{agrilife2024, WangEPICReview}. We include the EPIC crop model mathematical formulation below for completeness~\cite{singh}. The phenological development of crop $j$ is based on the daily heat unit accumulation,
\begin{align}
    \text{HU}^d = \max\left(0,  \frac{T_{\text{min}}^{d}+T_{\text{max}}^{d}}{2} - T_{\text{b},j} \right), \label{eqn: heat unit}
\end{align}
where $\text{HU}^d \in \mathbb{R}^{+}$ is the daily heat accumulation of day $d$ ($^\circ$C), $T_{\text{min}}^d$ and $T_{\text{max}}^d$ are the minimum and maximum temperatures of day $d$ ($^\circ$C), and $T_{\text{b},j}$ is the minimum temperature for growth to occur in crop~$j$ ($^\circ$C). 

The heat unit index $\text{HUI}^d$ of crop $j$ at day $d$ is used to track growth from planting ($\text{HUI}^{d=0}$) to the potential heat units $\text{PHU}_j$ needed for plant maturity,
\begin{align}
    \text{HUI}^d &= \frac{\sum_{i \leq d} \text{HU}^i}{\text{PHU}_j}.\label{eqn: HUI} 
\end{align}
Plant growth can be inhibited by a number of factors such as water, nutrient uptake, aeration, and temperature stress. Here, we consider the daily temperature stress in the crop stress factor $\text{REG}^d$,
\begin{align}
    \text{REG}^d = \sin{\left(\frac{\pi}{2} \left( \frac{T_{\text{G}}^d -T_{\text{b},j} }{T_{\text{o},j}-T_{\text{b},j}}\right) \right)}, \label{eqn: REG}
\end{align}
where $0\leq \text{REG}^d\leq 1$, $T_{\text{G}}^d$ is the average daily soil surface temperature ($^\circ$C), and $T_{\text{o},j}$ is the optimal crop growth temperature ($^\circ$C). It should also be noted that plant growth does not occur on days when the average daily temperature is 50\% higher than optimal crop growth temperature, i.e., $T_{\text{G}}^d > 1.5 T_{\text{o},j}$~\cite{singh}. In this work, we assume that the soil surface temperature is the same as the average daily temperature. 

EPIC then estimates the LAI on day $d$, denoted $\text{LAI}^d$, as
\begin{align}
    \text{LAI}^d = &\sum_{i\leq d} \left( \text{HUF}^i - \text{HUF}^{i-1} \right) \cdot \overline{\text{LAI}_j}\notag\\ &\cdot \left( 1-\text{exp}(5\cdot(\text{LAI}^{i-1} - \overline{\text{LAI}}_j)) \right)\cdot \sqrt{\text{REG}^i}, \label{eqn: LAI}\\
    \text{HUF}^d = & \frac{\text{HUI}^d}{\text{HUI}^d + \text{exp}(\text{ah}_{j1} - \text{ah}_{j2} \cdot \text{HUI}^d)}, \label{eqn: HUF}
\end{align}
where $\overline{\text{LAI}_j}$ is the maximum LAI of crop $j$, $\text{HUF}^d$ is the heat unit factor of day $d$, and parameters $\text{ah}_{j1}$ and $\text{ah}_{j2}$ are specific to crop $j$'s leaf area development curve. The LAI starts at zero and increases exponentially during early growth until it plateaus and then eventually declines. Assuming that the crop is harvested before it starts declining, we do not explicitly include the leaf decline equations that are for the period of time after the LAI plateaus. 

In Section~\ref{subsection: shading analysis}, we determined the field PAR given the shading from PV panels. However, not all PAR that reaches the field will reach the crops. This depends on the LAI, where earlier in the growing season, there is less leaf area for the interception of solar radiation. To account for this, the EPIC crop growth model calculates the PAR intercepted by the crops as a function of the LAI ($\text{Wh}/\text{m}^2$),
\begin{align}
    \text{PAR}_\text{crop}^d &= \sum_{t\in \mathcal{T}_d} \text{PAR}_\text{field}^t {\cdot (1 - \text{exp}({-0.65\cdot\text{LAI}^d})) \cdot \Delta T},\label{eqn: PAR crop SF}
\end{align}
where $\mathcal{T}_d$ is the set of time steps $t$ within day $d$. 

The biomass $B^d$ (t/ha) and final dry matter yield $Y$ (t/ha) are then 
\begin{align}
    B^d &= \sum_{i\leq d} 0.001 \cdot \text{BE}_j \cdot \text{PAR}_\text{crop}^i \cdot \text{REG}^i,\label{eqn: biomass}\\
    Y &= \text{HI}_j \cdot B^{d=|\mathcal{D}|}, \label{eqn: yield}
\end{align}
where $\text{BE}_j$ is the crop parameter for converting energy to biomass and $\text{HI}_j$ is the harvest index of crop $j$, which incorporates the fraction of the crop that is above ground. It should be noted that the yield is a single value at the end of the growing period. The crop LER is then 
\begin{align}
    \text{LER}_\text{crop} &= \frac{Y}{Y_\text{crop}},\label{eqn: LER crop}
\end{align}
where parameter $Y_\text{crop}$ is the yield under the crop-only case.

\section{MPC framework} \label{section: Optimization}
In this section, we present the MPC optimization formulation. MPC is well-suited for agrivoltaic PV panel scheduling under model uncertainty (e.g., imperfect crop modeling) and parameter uncertainty (e.g., irradiance and temperature modeling). Because trajectory predictions lose accuracy over the horizon as the system diverges from the expected state, MPC recalculates the optimal trajectory each time step but only implements the immediate control action. The optimization determines the PV panel operation over the entire optimization horizon, implementing the control decision for the current time step and then re-solving the problem for the remaining horizon as new measurements and forecasts of irradiance and temperature become available, as illustrated in Fig.~\ref{fig: block diagram}. The MPC block in Fig.~\ref{fig: block diagram} represents the optimization problem that is solved every period with updated measurements and forecasts which can be written as 
\begin{maxi*}|s|[2] 
{\mathbf{x}   }
{ \omega\cdot \text{LER}_\text{pv} + (1-\omega)\cdot \text{LER}_\text{crop}} {}{} \tag*{{(P1)}}
\addConstraint{\!\!\!\text{Power/Shading Model: }\eqref{eqn: PAR GHI ratio}\text{-}\eqref{eqn: incidence angle}, \eqref{eqn: ST tilt}\text{-}\eqref{eqn: power adjust}, \eqref{eqn: PAR field SF} \; &\forall\, t...T}{}
\addConstraint{\!\!\!\text{Crop Model: }\eqref{eqn: heat unit}\text{-}\eqref{eqn: biomass} \quad  &\forall\, d...D}{}
\addConstraint{\!\!\!\text{PV LER: }\eqref{eqn: PV cost}, \eqref{eqn: LER pv}}{}
\addConstraint{\!\!\!\text{Crop LER: }\eqref{eqn: yield}, \eqref{eqn: LER crop},}{}
\end{maxi*}
where $\omega$ is a weighting coefficient between 0 and 1 that indicates the relative priority of PV outcomes ($\omega=1$) and crop outcomes ($\omega =0$). In (P1), we solve for the PV panel adjustments away from the sun-tracking position subject to the crop growth model, shading analysis, and PV power production constraints over the entire growing season. The decision variables in $\mathbf{x}$ are $\text{LER}_\text{pv}$, $\text{LER}_\text{crop}$, $\delta I_\text{db}^t$, $\delta I_\text{diff}^t$, $\delta P^t$, $\delta\phi_\text{pv}^t$, $\delta\Sigma_\text{pv}^t$, $\phi_\text{pv}^t$, $\Sigma_\text{pv}^t$, $\theta_\text{s-pv}^t$, $\text{PAR}_\text{field}^t$, $B^d$, $Y$, and $\text{PAR}_\text{crop}^d$. The objective maximizes the weighted combination of $\text{LER}_\text{pv}$ and $\text{LER}_\text{crop}$, allowing us to examine the trade-offs between PV energy revenue and crop yield. Alternatively, the objective could be formulated to maximize total economic value (e.g., in dollars), where crop revenue can be calculated given the final yield and market prices. 

The optimization formulation (P1) is  nonconvex  due to the angle of incidence \eqref{eqn: incidence angle}, irradiance deviations \eqref{eqn: db irradiance adjust}-\eqref{eqn: diff irradiance adjust}, and shading factor function in \eqref{eqn: PAR crop SF}. Because  nonconvex problems can be difficult to solve, we apply variable redefinitions, convex relaxations, and an approximation of the shading factor function to reformulate the problem as a convex program. At each time step, we evaluate our optimal PV tilt angle within the full power, shading, and crop model. The following subsections describe our approach. Its performance relative to the original, nonconvex formulation is evaluated in Section~\ref{subsection: computational performance}.

\subsection{Shading factor function approximation}\label{subsection: sf approx}
We first evaluate the relationship between the shading factor and  PV panel position. To do this, we simulate the shading pattern resulting from adjustments to the PV panel tilt and azimuth away from the sun-tracking position. Specifically, one degree of freedom (i.e., PV azimuth or tilt) is fixed while the other is adjusted. The resulting change in shading factor relative to the sun-tracking algorithm is computed over the full range of feasible angle adjustments, with a resolution of $1^\circ$. For the case studies presented in Section~\ref{section: Case Study}, this process is repeated for each one-hour time step over the growing season. 

We observe that varying the azimuth has little impact on the shading factor. This is because adjusting the azimuth corresponds to rotating the panel around the zenith-axis. For example, when the panel tilt is zero, changing the azimuth angle does not alter the shading pattern within the field, except for cases where the shadows extend beyond the field boundary. Consequently, in our formulation, we do not adjust the PV panel azimuth away from the sun-tracking trajectory. 

\begin{figure}
    \centering
	\includegraphics[width=0.8\linewidth]{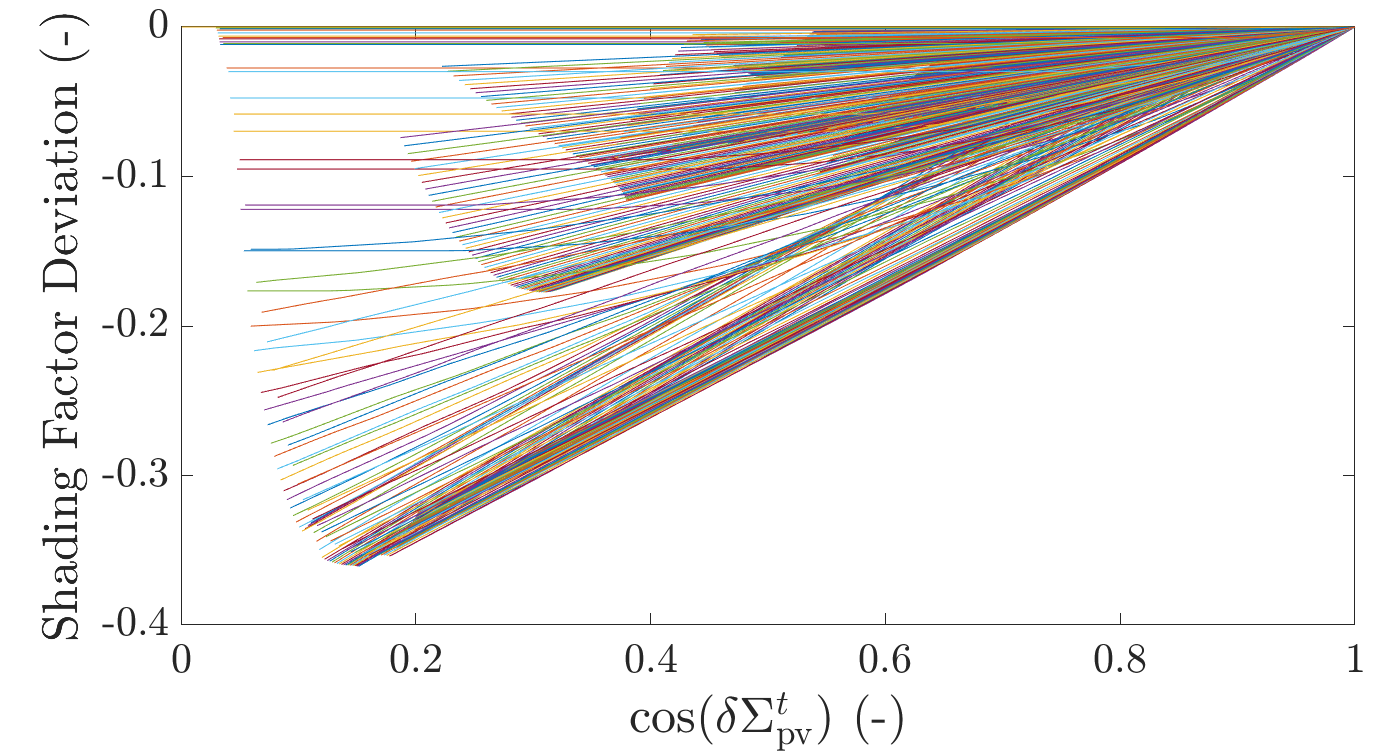}
	\caption{Changes in shading factor as a function of the cosine of tilt angle adjustments away from the sun-tracking trajectory \eqref{eqn: ST tilt}-\eqref{eqn: ST azimuth}. Each line corresponds to a one-hour time step over the growing season, evaluated with $1^\circ$ resolution. The full range of feasible tilt angle adjustments were considered at each time step, i.e.,  $\Sigma_\text{pv}^t \in [\Sigma_\text{pv,st}^t  \pm90^\circ  \text{ s.t. } \eqref{eqn: PV tilt limit}]$.  }
	\label{fig: linear shading deviations}       
\end{figure}

\begin{figure}
    \centering
	\includegraphics[width=\linewidth]{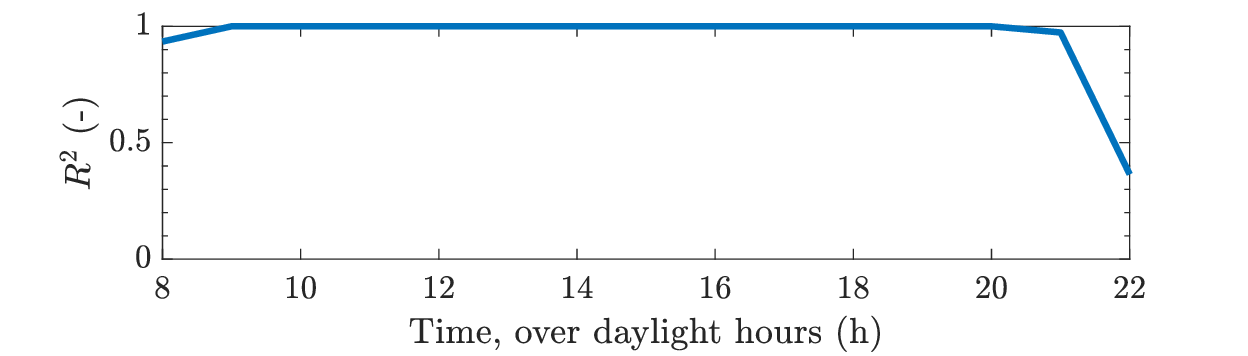}
	\caption{Hourly $R^2$ values of the linear approximation \eqref{eqn: SF linear}, averaged over the growing season. }
	\label{fig: Rsquared}       
\end{figure}

Alternatively, we observe that tilt adjustments away from the sun-tracking trajectory always reduce the shading factor. In Fig.~\ref{fig: linear shading deviations}, we observe that the change in the shading factor from the sun-tracking algorithm at every time step $t$ exhibits a strong linear relationship with the cosine of the tilt angle adjustment, $\cos(\delta\Sigma_\text{pv}^t)$. Motivated by this, we approximate the shading factor at time $t$ with an affine function of $\cos(\delta\Sigma_\text{pv}^t)$, i.e.,
\begin{align}
    S_\text{F}^t  =  g_1^t \cdot \cos{(\delta\Sigma_\text{pv}^t)} + g_2^t ,\label{eqn: SF linear}
\end{align}
where $g_1^t$ and $g_2^t$ are parameters of the affine approximation. The residuals between the estimated linear function and the actual shading factor deviations were small, generally less than $10^{-4}$. Fig.~\ref{fig: Rsquared} shows the hourly time-of-day $R^2$ values of the best-fit linear approximation, averaged over the entire growing season. We found that the $R^2$ values are equal to one for all time periods except for those within the first and last daylight hours. The $R^2$ values less than one are primarily due to the low solar altitude when the sun rises and sets. For instance, in hour 22, the sun is above the horizon for only seven days of the 60-day growing season in our case study set up, with very small solar altitude angles between $1.7^\circ$ and $1.9^\circ$. Under these conditions, shadows are often projected outside of the crop field boundaries, leading to flattened parts of the shading factor curve. However, these approximation errors have limited impact on overall results because PAR intensity during these periods is much lower than  during peak daylight hours and it only affects a small number of time steps. 
\subsection{Variable redefinitions} \label{subsection: variable redef}
To improve the computational tractability, we reformulate and relax (P1) to obtain a convex problem, which is an approach commonly employed in the optimal control space (e.g., \cite{Malyuta2022}). We introduce the variables 
\begin{align*}
    x^t:=\cos(\delta\Sigma_\text{pv}^t), \quad y^t:=\sin(\delta\Sigma_\text{pv}^t),
\end{align*}
and apply trigonometric identities
\begin{align*}
    \sin(x + y) = \sin(x)\cos(y) + \cos(x)\sin(y),\\
    \cos(x + y) = \cos(x)\cos(y) - \sin(x)\sin(y),
\end{align*}
to rewrite the nonconvex constraints as linear functions of $x^t$ and $y^t$. Specifically, the irradiance deviations \eqref{eqn: db irradiance adjust}-\eqref{eqn: diff irradiance adjust}, the PAR received by the crops \eqref{eqn: PAR crop SF}, and the PV tilt limits \eqref{eqn: PV tilt limit} can be expressed as 
\begin{align}
    \delta I_\text{db}^t  &= \text{DNI}^t x^t -\text{DNI}^t, \label{eqn: db irradiance adjust linear}\\
    \delta I_\text{diff}^t &= b_1^t x^t + b_2^t y^t - b_1^t,  \label{eqn: diff irradiance adjust linear}\\
    \text{PAR}_\text{field}^t &= c_1^t x^t + c_2^t,  \label{eqn: PAR crop linear}\\
    0\leq \,&d_1^t x^t + d_2^t y^t \leq 1,  \label{eqn: PV tilt limits redefined}
\end{align}
where parameters $b_1^t$, $b_2^t$, $c_1^t$, $c_2^t$, $d_1^t$, and $d_2^t$ depend on the solar irradiance and position
\begin{align*}
    b_1^t &:= \frac{1}{2}\text{DHI}^t \cos(90^\circ-\beta_\text{s}^t),\\
    b_2^t &:= -\frac{1}{2}\text{DHI}^t \sin(90^\circ-\beta_\text{s}^t),\\
    c_1^t &:= -g_1^t\cdot \text{DNI}^t \alpha,\\
    c_2^t &:= \alpha \cdot( \text{DNI}^t - g_2^t \text{DNI}^t +  \text{DHI}^t),\\
    d_1^t &:= \sin(90^\circ-\beta_\text{s}^t) ,\\
    d_2^t &:= \cos(90^\circ-\beta_\text{s}^t).
\end{align*}
The derivation of this approach is outlined in the appendix of~\cite{StuhlmacherHICSS2024}, with minor changes to the PAR field calculation to separate the direct beam and diffuse PAR contribution.

When replacing $\cos{(\delta\Sigma_\text{pv}^t)}$ and $\sin{(\delta\Sigma_\text{pv}^t)}$ with $x^t$ and $y^t$, we need to enforce
\begin{align}
    (x^t)^2 + (y^t)^2 =1 , \label{eqn: quadratic equality}
\end{align}
to ensure that a unique value of $\delta\Sigma_\text{PV}^t$ can be recovered. We relax \eqref{eqn: quadratic equality} to a second order cone constraint 
\begin{align}
    (x^t)^2 + (y^t)^2 \leq 1 ,\label{eqn: quadratic relaxation}
\end{align}
and evaluate the exactness of this convex relaxation in Section~\ref{subsection: computational performance}.

\subsection{Convex reformulation} \label{subsection: convex reformulation}
Applying the approximations and relaxations described in Sections~\ref{subsection: sf approx} and \ref{subsection: variable redef}, the optimization problem (P1) can be reformulated as a convex SOCP  
\begin{maxi*}|s|[2] 
{\mathbf{x}}
{ \omega \cdot \text{LER}_\text{pv} + (1-\omega)\cdot \text{LER}_\text{crop}} {}{} \tag*{{(P2)}}
\addConstraint{\!\!\text{Power/Shading Model: }\eqref{eqn: power adjust}, \eqref{eqn: db irradiance adjust linear}\text{-}\eqref{eqn: PV tilt limits redefined}, \eqref{eqn: quadratic relaxation} \;  &\forall\, t=1...T}{}
\addConstraint{\!\!\text{Crop Model: }\eqref{eqn: heat unit}\text{-}\eqref{eqn: biomass} \quad  &\forall\, d...D}{}
\addConstraint{\!\!\text{PV LER: }\eqref{eqn: PV cost}, \eqref{eqn: LER pv} }{}
\addConstraint{\!\!\text{Crop LER: } \eqref{eqn: yield}\text{-}\eqref{eqn: LER crop}. }{}
\end{maxi*}
The decision variables in $\mathbf{x}$ are $\text{LER}_\text{pv}$, $\text{LER}_\text{crop}$, $x^t$, $y^t$, $\delta I_\text{db}^t$, $\delta I_\text{diff}^t$, $\delta P^t$, $\text{PAR}_\text{field}^t$, $\text{PAR}_\text{crop}^d$, $B^d$, and $Y$. If \eqref{eqn: quadratic equality} holds for the optimal $x^t$ and $y^t$, then \eqref{eqn: quadratic relaxation} is exact and the tilt adjustment  $\delta \Sigma_\text{pv}^t$ can be recovered by solving $\delta \Sigma_\text{pv} = \sin^{-1}(y^t)$. Otherwise, we recover a more conservative tilt deviation by selecting the larger magnitude tilt deviation from $x^t$ and $y^t$ separately. This results in a tilt adjustment that may allow more PAR to reach the crops and produce less PV power than expected. This behavior is examined in Section~\ref{subsection: computational performance}.

\section{Case studies} \label{section: Case Study}
We evaluate our approach with a simulated agrivoltaic system. We use the same parameters as~\cite{StuhlmacherHICSS2024}, where we now consider the inputs and outcomes over an entire growing season, incorporate the EPIC crop growth model for lettuce, and consider forecast errors and updates within an MPC framework. 

\subsection{Set up} \label{subsection: Set Up}
We evaluate the performance of a synthetic agrivoltaic system located in Ann Arbor, Michigan with 280 PV panels, as described in~\cite{StuhlmacherHICSS2024} and the corresponding publicly available source code~\cite{Githubcode}. The set up consists of 7 rows, with 40 PV panels per row. We also use the same time-varying electricity price as~\cite{StuhlmacherHICSS2024}, which is repeated every day.

We use historical solar irradiance and temperature data. We pull $\text{DHI}^t$, $\text{DNI}^t$, and $T^t$ values from the National Solar Radiation Database (NSRDB)~\cite{NREL_SPA}. We consider lettuce since it is a shade tolerant crop commonly analyzed in agrivoltaics research (e.g., \cite{dinesh_2016, StuhlmacherHICSS2024, marrou_2013, elamri_2018, mazzeo2025optimizing}), where all crop parameters are pulled from WinEPIC~\cite{agrilife2024}. The growing season horizon is 60 days, where we pull hourly NSRDB historical data from July and August 2023. As a result, $T = 1440$ and $D=60$. The \texttt{get\_solarposition} function in the Python package pvlib is used to calculate the solar azimuth $\phi_\text{s}^t$ and altitude $\beta_\text{s}^t$~\cite{holmgren2018pvlib}. 

\begin{figure}
\centering \includegraphics[width=0.8\linewidth]{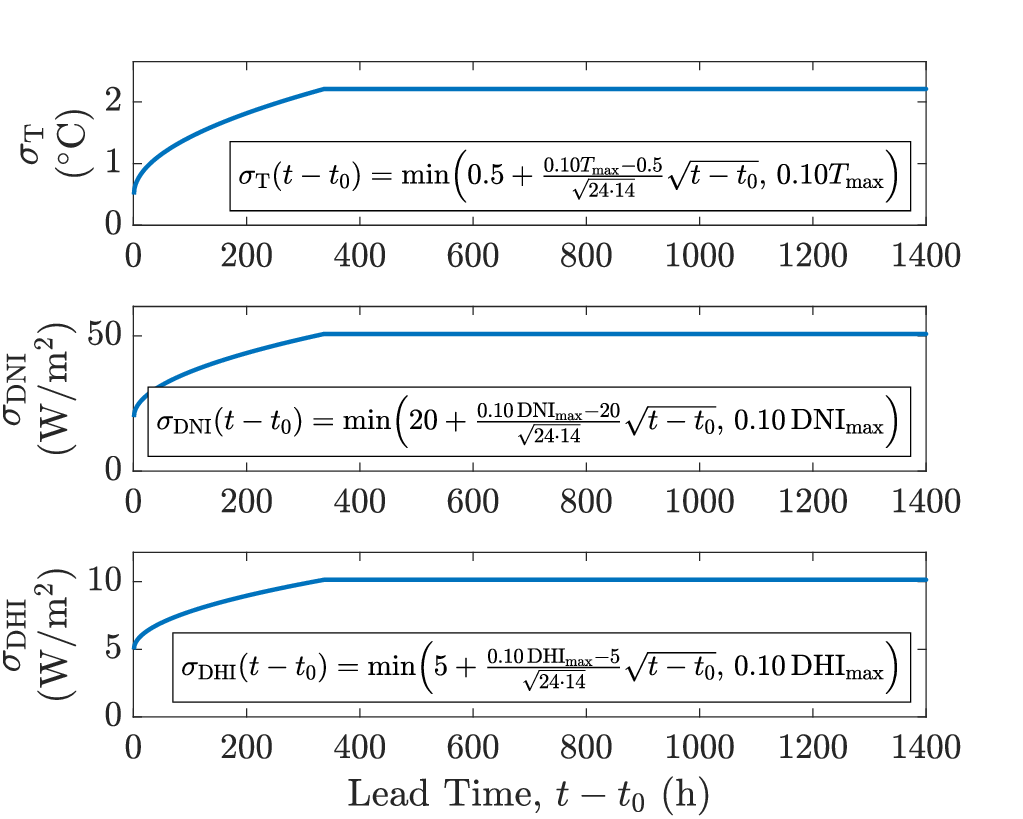} \caption{Standard deviation for temperature, DNI, and DHI Gaussian noise as a function of the forecast lead time $t-t_0$. The standard deviation increases with the forecast lead time and is capped at two weeks. In this example, the maximum standard deviation is capped at 10\% of the expected range of each variable. } \label{fig: noise variance} \end{figure}

For realism, we use imperfect irradiance and temperature forecast profiles, which are updated at each time step. To generate the updated forecast profiles, we use an autoregressive linear model AR(1), that updates the forecast based on the previous forecast error. For example, the temperature forecast used at time $t_0$ for a future time $t$ would be
\begin{align}
    T^{t} = \mu_\text{T}^{t} + \gamma \cdot (\mu_\text{T}^{t-1} - T^{t-1} ) + \epsilon_\text{T}^{t}(t_0),
\end{align}
where $\mu_\text{T}^{t}$ is the actual temperature at time $t$, $\gamma$ is the autocorrelation parameter, and $\epsilon_\text{T}^t(t_0)$ is Gaussian process noise with zero mean and a variance that increases with the forecast lead time $t-t_0$, i.e., 
$\epsilon_\text{T}^{t}(t_0) \sim \mathcal N(0, \sigma_\text{T}({t}-t_0)^2)$. The temperature noise standard deviation function $\sigma_\text{T}({t}-t_0)$ is increasing with respect to the forecast lead time; specifically, it is dependent on the square root of the lead time and is capped at a specified maximum at two weeks out. An analogous formulation is used for DNI and DHI. To explore different levels of uncertainty, we set $\gamma$ to 0.8 and scale the maximum value of the noise standard deviation relative to the expected range of each parameter. Fig.~\ref{fig: noise variance} shows the standard deviation functions for temperature, DNI, and DHI when the maximum standard deviation is set to 10\% of the expected range. We solve (P2) with the Gurobi solver \cite{Gurobi} using the JuMP package in Julia. We vary $\omega$ to investigate the trade-offs in crop and PV outcomes.

\subsection{Crop-PV trade-offs and land equivalent ratio (LER)}
\label{subsection: Pareto front tradeoffs and LER}

\begin{figure}
    \centering
	\includegraphics[width=\linewidth]{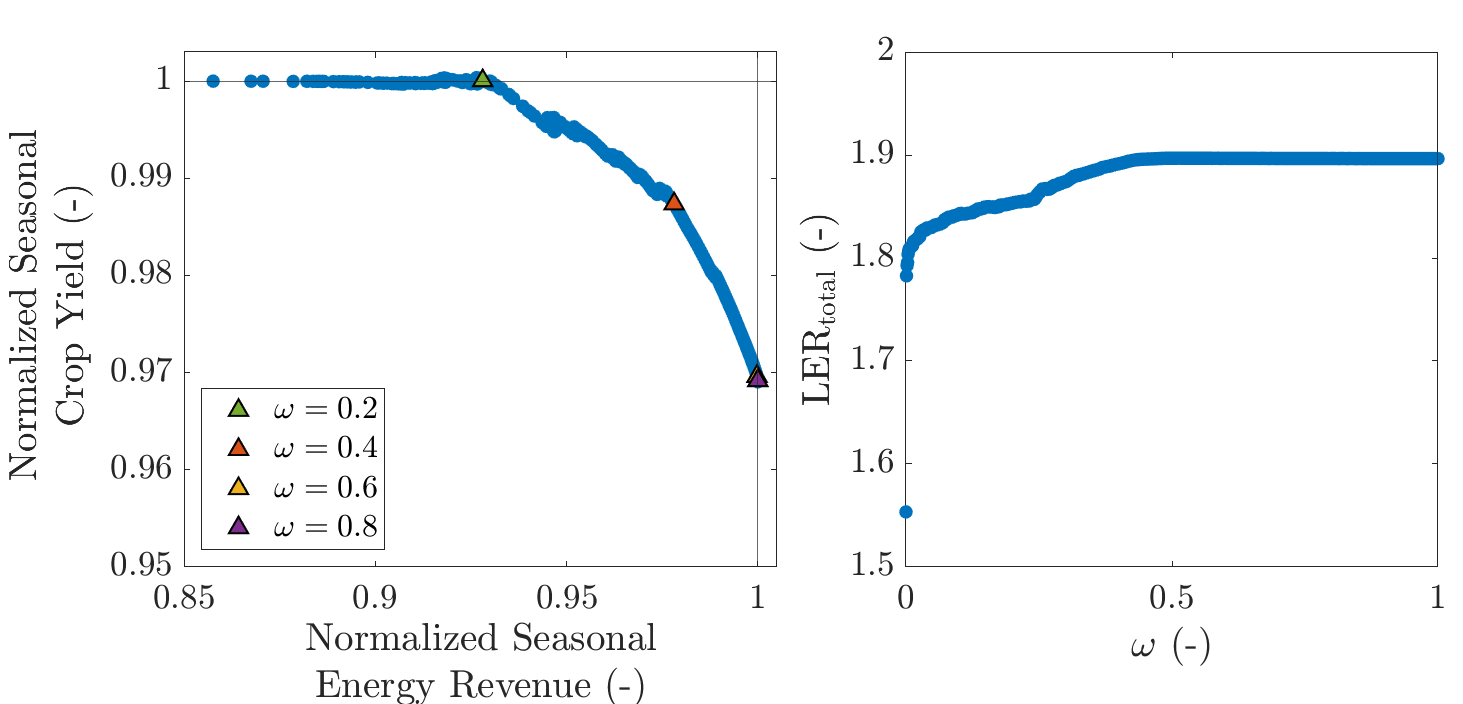}
	\caption{(Left) Trade-off between seasonal crop yield and energy revenue as the weighting coefficient $\omega$ varies. Crop yield and energy revenue are normalized relative to the values at $\omega = 0$ and $\omega=1$, respectively. (Right) Total land equivalent ratio $\text{LER}_\text{total}$ as a function of weighting coefficient $\omega$. }
	\label{fig: pareto front}       
\end{figure}

We first investigate the impact of the weighting coefficient~$\omega$, which specifies the relative importance of crop yield and power production in the objective function. To isolate the impact of $\omega$ on the optimal solution, we solve (P2) over the entire growing season assuming that the temperature and irradiance are known in advance and evaluate our optimal solution within the full crop, shading, and PV model to get the actual crop yield and energy revenue. We vary the weighting coefficient $\omega$ between zero (i.e., maximizing crop yield only) and one (i.e., maximizing energy production only), with increments of 0.001. The left plot of Fig.~\ref{fig: pareto front} illustrates the resulting trade-off between the normalized seasonal crop yield and the normalized seasonal energy revenue. As power production is prioritized, seasonal crop yield decreases. 

The Pareto front and associated total LER values computed with~\eqref{eqn: LER total} visualize these trade-offs. As shown in the right plot of Fig.~\ref{fig: pareto front}, the LER increases for small values of $\omega$ and then remains relatively constant for $\omega$ values greater than 0.4, where the highest $\text{LER}_\text{total}$ occurs at $\omega$ values between 0.492 and 0.562. This corresponds to a total LER of 1.897. These results can inform the selection or tuning of $\omega$ for different agrivoltaic configurations or operational priorities. Because the total LER varies only slightly for $\omega > 0.4$, the solution is relatively insensitive to small changes in $\omega$, indicating that precise tuning of this parameter is not critical.

\subsection{Computational performance} \label{subsection: computational performance}
\begin{figure}
    \centering
	\includegraphics[width=0.8\linewidth]{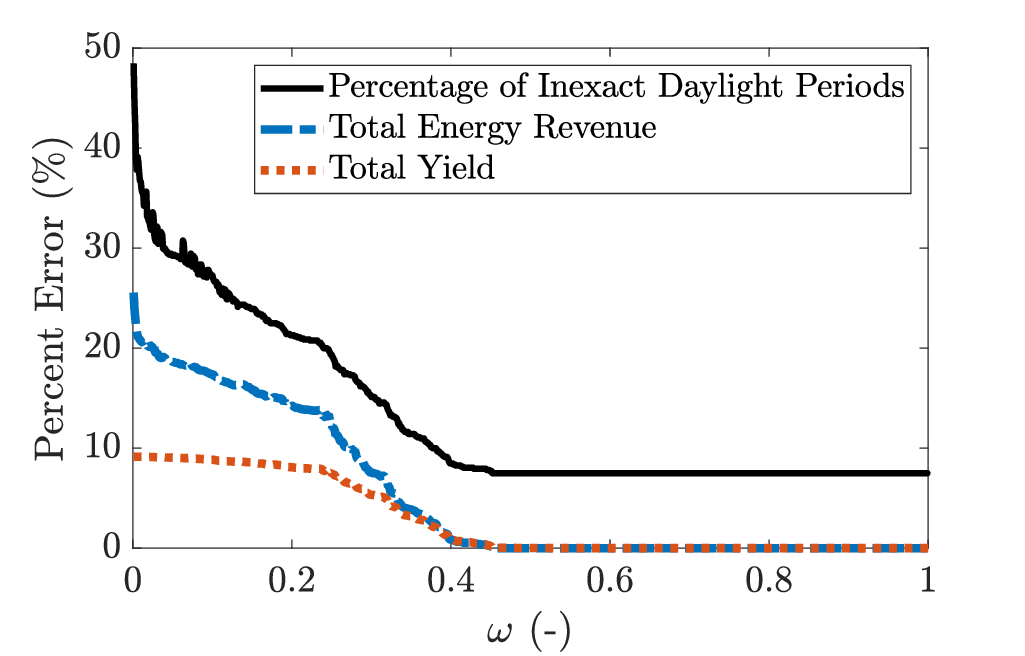}
    \vspace{-0.2cm}
	\caption{Percent error of crop yield and energy revenue as function of $\omega$.  Percent error evaluates the differences between modeling outcomes from (P2) and the realized outcomes computed using the original nonconvex model. The blue dashed and red dotted lines show the percent error in energy revenue and crop yield, respectively. The solid black line shows the percentage of daylight periods within the growing season for which the relaxation~\eqref{eqn: quadratic relaxation} in (P2) is not exact.}
	\label{fig: percent error}     
    \vspace{-0.1cm}
\end{figure}

We next evaluate the accuracy of our (P2) formulation, focusing on the impact of the convex relaxation~\eqref{eqn: quadratic relaxation} and the shading factor approximation~\eqref{eqn: SF linear}. Prior work in~\cite{StuhlmacherHICSS2024} found that the shading factor approximation closely matched the actual shading factor, with slightly larger deviations during the first and last daylight hours. This behavior is also reflected by the average hourly $R^2$ values being less than one during those periods in Fig.~\ref{fig: Rsquared}. 

We find that the relaxation is not exact for all time periods and values of $\omega$. The solid black line in Fig.~\ref{fig: percent error} shows the percentage of daylight periods over the entire growing season where the relaxation is inexact for different values of $\omega$, ranging from around 50\% to 7.5\%. Exactness improves as $\omega$ increases and then plateaus for $\omega>0.4$. This behavior likely occurs because for small values of $\omega$, energy revenue is a minor contributor to the objective, making PV-related decision variables less influential and impacting the tightness of the relaxation. 

When the solution is inexact, we recover the PV tilt angle by choosing the largest magnitude of $\delta \Sigma_\text{pv}^t$ from $x^t$ and $y^t$. Under this strategy, the PV panels may be adjusted further away from the sun-tracking position, reducing the capture of direct beam irradiance while allowing more sunlight through to the crops. As a result, the actual crop biomass and yield may be higher than what is estimated in (P2) while the total energy production may be lower. However, selecting the larger tilt angle deviation may sometimes lead to $\Sigma_\text{pv}^t$ values outside of the physical range we chose in our case study, i.e., $0 \leq \Sigma_\text{pv}^t \leq 90^\circ$. When this occurs, we force the tilt angle to be within its limits, which can cause the expected bias (an overestimate of power and underestimation of yield) to not always hold. 

To quantify the difference between the expected and actual agrivoltaic outcomes, we compute the percent error. This is done by plugging the optimal decisions $\delta \Sigma_\text{pv}^t$ from (P2) into the full power, shading, and crop model to compute the actual energy revenue and crop yield. The percent error formula is used to compare the actual yield and energy revenue from the original nonconvex model with the expected values from (P2) in Fig.~\ref{fig: percent error}. The dotted red and dashed blue lines in Fig.~\ref{fig: percent error} show the percent error of yield and energy revenue as we vary $\omega$. We observe that the percent error is higher when there are more inexact solutions ($\omega < 0.4$). For $\omega$ values greater than 0.4, the percent error becomes negligible for all cases, approximately 0.01\%, even though the relaxation remains inexact in about 7.5\% of the daylight time periods. As discussed in \cite{StuhlmacherHICSS2024} and Section~\ref{subsection: sf approx}, these inexact periods and shading factor approximation errors generally occur in the first and last daylight hours, where there is less sunlight in the morning and evening hours compared with midday.  When we factor in the irradiance magnitudes, the approximation errors are less significant in the first and last
hours. Consequently, their impact on season-long outcomes of yield and energy revenue is minimal.

\subsection{MPC performance} 
Last, we evaluate the performance of the MPC approach and the impact of forecast uncertainty. Table~\ref{table: mpc comparison} compares the final LER values when $\omega=0.5$ under a perfect forecast and under the MPC approach with increasing forecast uncertainty. In the MPC cases, the standard deviation of the Gaussian process noise is varied from 5\% to 15\% of the expected range of DHI, DNI, and temperature.  As expected, increasing forecast uncertainty leads to  a lower total LER value, primarily due to an decrease in PV energy revenue. This occurs because the forecast uncertainty impacts each subsequent step, resulting in solutions that are suboptimal relative to the perfect forecast case. However, we find that the total LER achieved by the MPC approach remains close to that obtained under perfect forecasts, indicating that the controller performs well under uncertainty. Importantly, even with forecast uncertainty, the MPC solutions with $\omega=0.5$ still yield a higher total LER than the perfect forecast case where we only maximize crop performance ($\omega=0$).

\begin{table}
\centering
\caption{LER comparison of perfect forecast and MPC with varying forecast noise ($\omega = 0.5$)}
\vspace{-0.1cm}
\label{table: mpc comparison}
\begin{tabular}{lrrr}
\hline \bf Approach & \multicolumn{3}{c}{\bf LER (-) } \\ 
 & \bf Crop & \bf  PV & \bf  Total \\ \hline
Perfect Forecast & 0.899 &   0.998 & 1.897\\
MPC, $5\%$ & 0.896 &   0.959 & 1.855\\
MPC, $10\%$  & 0.896 &   0.957 & 1.853\\
MPC, $15\%$  & 0.896 &   0.955 & 1.851\\
\hline
\end{tabular}
\vspace{-0.1cm}
\end{table}

\section{Discussion} \label{section: Discussion}

Our results show that agrivoltaics can balance power production and crop needs through the active control of the dual-axis PV panels. By adjusting the PV tilt away from the sun-tracking trajectory, small reductions in energy production can lead to increases in crop yield, improving the overall LER.

While this paper presents power production and crop yield as a trade-off, there is significant work showing that the presence of solar PV panels above crops can, in some cases, increase crop yields due to the creation of micro-climates~\cite{marrou_2013} more conducive to crop growth \cite{barron-gafford_2019, thakur2025exploring}. We have heard anecdotes of panel shading affecting the behavior of crop pests, with one type of pest avoiding shaded areas resulting in a reduction of crop damage. Panels can also affect soil health including nutrients and aeration~\cite{luo_early_2024}.
Moreover, the presence of crops underneath panels could affect solar power capture, since factors such as temperature and wind speed affect panel efficiency~\cite{kaldellis_temperature_2014}. Active panel control could also support crop irrigation goals, mitigate crop water stress, and/or contribute to farmland water management, e.g., through rain diversion, crop protection from frost and snow~\cite{widmer2024agrivoltaics}, evaporation management~\cite{hassanpour2018remarkable, elamri_2018}, dew management, erosion management~\cite{verheijen_discussion_2023}, etc. Capturing these various effects requires extending our crop models and/or crop-panel interaction models, and capturing more real-time measurement data (e.g., crop data, panel data, climate data) within the controller. These are important directions for future work. Furthermore, it will be important to explore the accuracy of existing crop models under (irregular) shading and other factors that are unique to agrivoltaics systems, such as reflected irradiance for bifacial panels with respect to the crop growth cycle~\cite{sanchez2024spectral} and the presence of ground-mounted equipment within the farmed site. Extended or novel models may need to be developed.

\section{Conclusion} \label{section: Conclusion}
This paper investigates the optimal operation of agrivoltaic systems with the goal of better understanding the operational trade-offs between the PV system and the crops. We develop an approach that optimizes the combined value of crop production and PV energy revenue within an agrivoltaic system. To capture the different timescales of the crop and PV power models, we considered the operational control of agrivoltaic panels over the entire growing season. 
Specifically, we formulate an optimization problem subject to PV-crop shading relationships and crop growth dynamics while maximizing the value of crop yield and energy revenue generated by the PV panels. We solve for the PV tilt angle deviation away from the sun-tracking trajectory.  To account for temperature and irradiance uncertainty, we implement an MPC framework where we update our control decision at each time step given new measurements and updated forecasts. 

Through case studies, we demonstrate that the proposed approach can successfully adjust the PV panels away from the sun-tracking position to balance PV and crop needs. We identify the Pareto front trade-off between crop yield and PV energy revenue by varying the weighting parameter $\omega$ and quantify the operational value of agrivoltaics using LER. In all cases examined, we found that agrivoltaic operation can achieve an LER larger than one, indicating a greater land productivity compared to single-use cases. 
We further assess the accuracy of the convex relaxation and shading approximations, finding that the resulting errors in seasonal energy production and crop yield are negligible for $\omega$ values greater than 0.4, demonstrating that the proposed formulation provides both computational tractability and accurate agrivoltaic performance estimates. Finally, we evaluate the MPC performance under forecast uncertainty and find that the controller maintains performance close to the perfect forecast case and continues to outperform the crop-only optimization scenario. Future work will focus on incorporating additional agrivoltaic system interactions and quantifying the impact on power systems operation.

\bibliographystyle{IEEEtran}
\bibliography{references.bib}

\end{document}